# A Computational Study of Mixing Microchannel Flows

John Patrick Bennett[1], Chris H Wiggins[2],
[1]Manhasset High School, Manhasset NY 11030
[2]Department of Applied Physics and Applied Mathematics
Center for Computational Biology and Bioinformatics
Columbia University, New York NY 10027

July 15, 2003

## Abstract

Motivated by recent experimental advances [1, 2] in microfluidic mixers, we study the passive mixing and flow properties of a patterned microchannel by means of computational fluid dynamics (CFD). Such geometries overcome the low Reynolds number, high Péclet number boundaries to efficient mixing by creating a three-dimensional flow that yields chaotic trajectories for advected passive scalars. The flow seems to exploit an effective shearing mechanism, or "ditch mixing," rather than the alternating span-wise vorticity mechanism we anticipated. Further, we find superexponential mixing for a modification of the geometry. It is hoped that such CFD studies advance both the capabilities and the understanding of such micropatterned mixing flows.

### I. Introduction

Ever-improving technological advances in micro- and nano- fabrication have encouraged renewed interest in mixing through topological or chaotic flows. Particular attention has been paid to geometries which are sufficiently small that neither turbulent (owing to the low Reynolds number), nor active (owing to the difficulty in micromachining moveable or easily controlled parts) mechanisms are feasible. Examples of recent successful strategies include a "staggered herringbone" micromixer [1] in which two species were mixed well beyond the diffusion limit simply through patterning the bottom of a channel 200 microns in width. The geometry, illustrated in Figure 1A, features an alternating bas-relief pattern of obliquely-angled short and long grooves, meeting at a point which is off-center, designed to induce two separate flow regions of opposite span-wise vorticity.

This last particular example of passive micromixing begs computational study both to determine the role of chaotic or geometric mixing, as well as possible improvements, either in maximizing mixing or minimizing pressure drop, through choosing different patterns or different parameter values for the pattern used in [1]. There are as many design options as there are imaginable, experimentally-realizable patterns for the bottom of the channel. Using the computational fluid dynamics software Fluent$^{TM}$ (V6), we have undertaken such a study. Identical physical parameters were used as those in [1], and, after verifying computational results against experimental results using the published geometry, and studying several properties of the flow which are not experimentally accessible, we studied several variations of the pattern: (i) identical channel with no grooves, (ii) identical channel in which the short legs were removed, (iii) identical channel with double-depth grooves, (iv) identical channel with half-depth grooves, and (v) a nanomixer, with all lengths rescaled by a factor of $10^{-3}$ in order to eliminate conclusively finite Reynolds number effects as a candidate for the mixing mechanism utilized.

For each of these geometries, the principal quantitative measure of mixing is the standard deviation in concentration of a passive mixing species (scalar). This quantity varies between 0.5 for a completely unmixed flow and 0 for a completely mixed flow, and was found in [1] to decay exponentially along the length of the channel. We also study the pressure drop along the length of the channel, which limits the velocity and therefore the time necessary for effective mixing. Advantages of full CFD studies of micromixing include the ability to visualize species densities at any point in the volume (without optical challenges of confocal-microscopy), as well as quantities relevant to mixing, e.g., vorticity in the span-wise direction. We comment below on some unsuspected geometric features of this flow.

### II. Mixing

In all geometries studied, the material parameters are those used in [1], e.g., a density of 1.2 gm/cm$^3$, and a viscosity of 0.67 gm/cm s (cf., footnote 30 of [1]). For all but the nanomixer, length scales were those used in [1], e.g., a channel height of 69.7 μm and a channel width of 200 μm. In the experiment, grooves are 30.6 μm in depth, with a width of 50 μm. The velocity field is calculated by specifying a plug-flow input with a velocity of 0.2 cm/s, developing into a Poiseuille-flow with a pressure of 0 applied at the channel outlet plane, 1,516 μm downstream from the input. The corresponding Reynolds number ($Re = Ul/v$, with the channel width as the characteristic length) is $0.7 \times 10^{-2}$. Once calculated, this velocity field is then used as the input to a second numerical routine solving for the concentration of a diffusive passive scalar. The input condition (see Figure 1B) was that of the experiment, uniform in the vertical, with two values (0 and 1) separated by a vertical line at the center of the input (cf. figure 3A of [1]). Although Fluent$^{TM}$ allows one to set the diffusion constant arbitrarily low, and thus the Péclet number arbitrarily high, the grid used for the species solver is the same as the grid for the fluid solver (a structured, cooper mesh consisting of $\sim 7.5 \times 10^5$ nodes) and is expected grossly to under-resolve the interface between the two species. The diffusion constant input was that of the experiment, $D = 2 \times 10^{-8}$ cm$^2$/s, with a corresponding Péclet number of $2 \times 10^5$, but the numerical Péclet number is limited by the number of points in the simulated grid. Estimating the smallest possible numerically-resolvable diffusion constant as $\Delta x / U$, we determine the largest possible numerical Péclet number, in the case where the number of points used was $N = 7.5 \times 10^5$, to be $(UL / D) = (L / \Delta x) = (N)^{1/3} \approx 10^2$, in remarkable agreement with the computational results (cf. caption of Figure 7). To overcome this problem, we also studied individual particle trajectories (Figure 6), using Fluent™'s Particle Injection feature interpolating mean velocities to track velocities at arbitrary points in the flow volume.

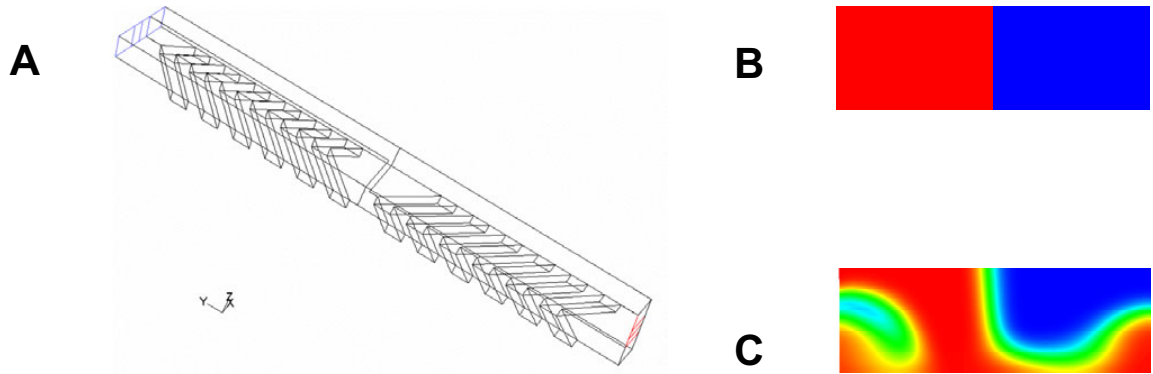

**Figure 1:** Geometry design (A) and first-cycle performance of the experimental micromixer (B, C). Input condition with scalar values of 0 and 1, showing complete segregation (B). Outlet condition after one cycle, displaying stretching and folding, as well as qualitative similarity to the experimental results (cf. Fig 3C of [1]) (C).

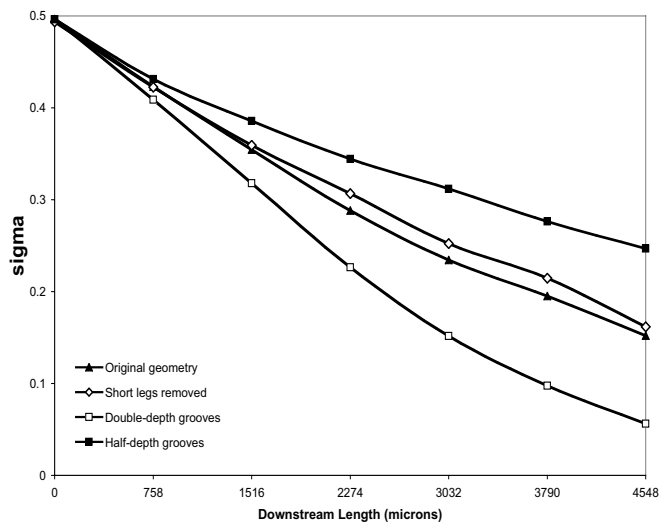
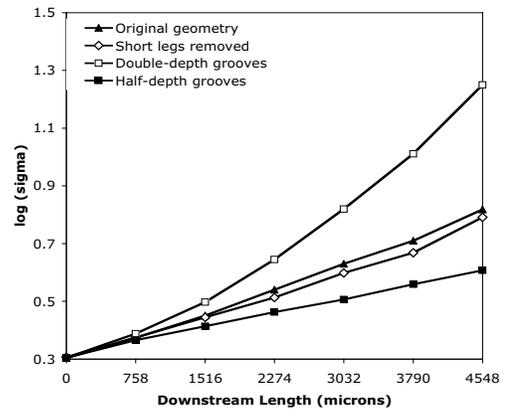
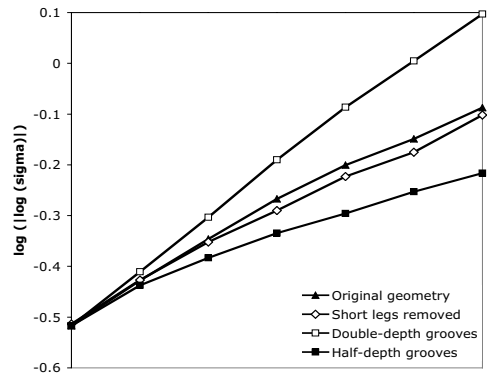

**Figure 2:** Standard deviation ($\sigma$), $\log(\sigma)$, and $\log(|\log(\sigma)|)$ of species density are plotted for several geometries. For all but the double-depth grooves, the dependence is exponential. The bottom right inset shows the $\log(|\log(\sigma)|)$ plot where superexponential decay is observed for the double-depth grooves.

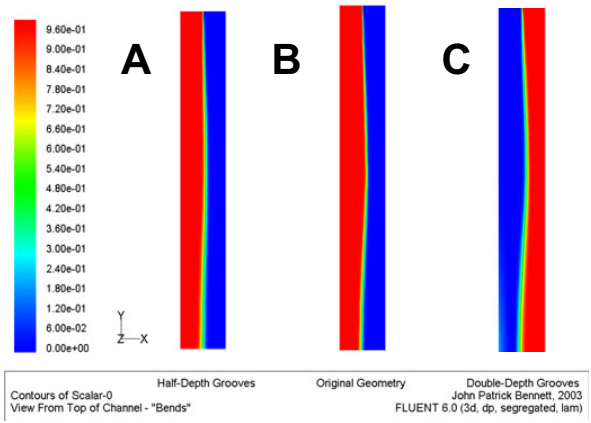

**Figure 3:** Contour maps of passive scalar near the top surface of the microchannel display limited effects of chaotic mixing on this region. Despite this limited disturbance, one can note a distinct increase in the amount of "bending" that occurs to the scalar interface as one goes from half-depth grooves (A) to original-depth grooves (B) and then to double-depth grooves (C).

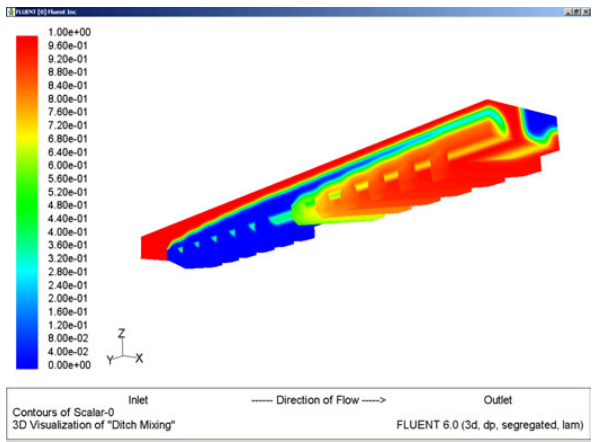

**Figure 4:** A three-dimensional visualization of the movement of species (scalar) inside the microchannel is a unique benefit of CFD. This visualization allows one to see the shuttling of species through the grooves and the development of flow patterns we term "ditch mixing."

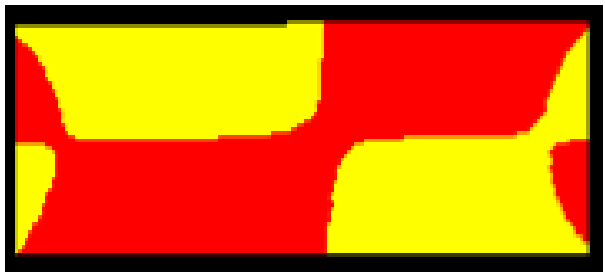

**Figure 5:** Plot of positive (red) and negative (yellow) regions of span-wise vorticity. This image displays, instead of the expected two-cell pattern, a more complicated 4-cell pattern.

Despite the under-resolution of the passive scalar field, images from Fluent™ look strikingly similar to those of the experiment (see Figure 1C). As in the experiment, we study the standard deviation as a function of length downstream. A plot of σ as a function of location shows nearly exponential decay for all but one of the conditions (cf. Figure 2). We note several interesting features: (i) the half-depth mixer has poorer mixing than the published geometry; (ii) the nanomixer, with identical shape but all lengths scaled by a factor of $10^{-3}$, has identical mixing properties, eliminating finite Reynolds number effects as a candidate mixing mechanism, and (iii) the double-depth geometry displays superexponential mixing, appearing as a straight line on a log(|log (σ)|) vs. distance plot (see inset, Figure 2). Superexponential decay of the variance is a consequence of the stretching of fluid elements by chaotic processes [3], and it has been recognized recently [4] that a phase of superexponential decay can exist if the stretching action of the flow is strong enough. These differing mixing abilities are also evident from the top view of the channel (see Figure 3). It is clear that the interface at the top of the channel is relatively unaffected by the mixing in the original (Figure 3B) and half-depth cases (Figure 3A), but bends appreciably for the double-depth case (Figure 3C).

### III. Ditch Mixing

The underlying mixing mechanisms are elucidated by visualizing the passive scalar concentration from below, i.e., in the grooves themselves. This is an option unavailable in the experiment, as the confocal-microscopy permits visualization only in a planar slice of the center 50% of the channel [1]. A curious feature in Figures 4 and 6 is the apparent use of the grooves to shuttle species from one side of the channel to the other, an effective shearing mechanism we term "ditch mixing." This differs from the mechanism one might imagine of alternating vorticities, which create a chaotic flow in the manner of blinking vortex Eulerian flows. Motivated by this observation, we also plot the vorticity in the span-wise direction. While we expected to see a two-color image of negative vorticity on the right and positive vorticity on the left, we instead see a much more complicated structure, with roughly four major discernable regions with four minor regions (Figure 5). This observation persists, although the shapes of the image change, as one advances along the length of the channel (a QuickTime movie illustrating this feature is downloadable at www.columbia.edu/~chw2). Ditch mixing suggests that the short legs of the pattern are inessential and may be removed without significant deleterious effect on the mixing. Numerical studies of this geometry (cf. Figure 2), which suggest a flow featuring vorticity of unique but alternating sign, confirm this intuition: the ditches facilitate transport via a combination of alternating shears, and the mixing is almost as effective as in the original geometry. As further illustration, we investigated particle trajectories in the original geometry, beginning with three rows of passive scalars as in Figure 6. Note how several yellow traces dip into the grooves and are transported to the other side.

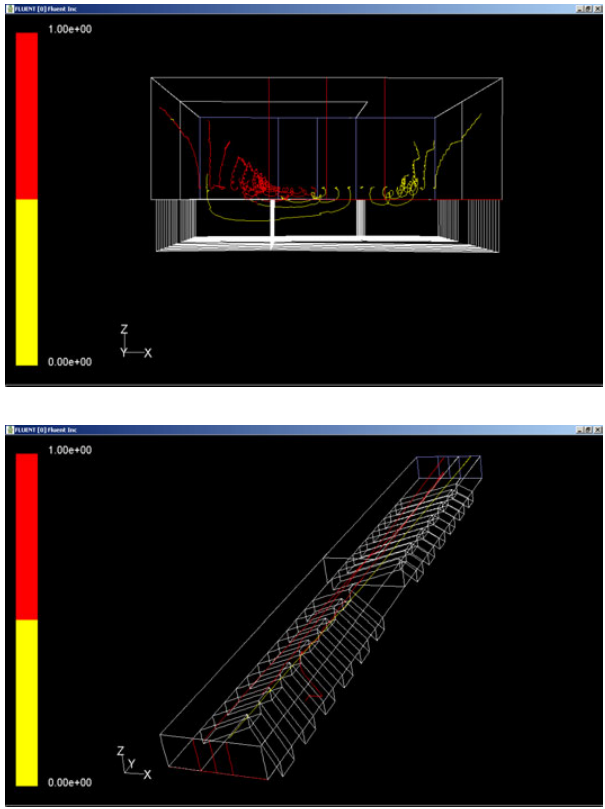

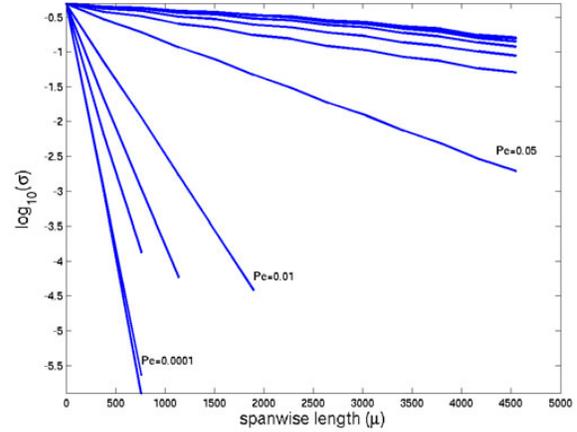

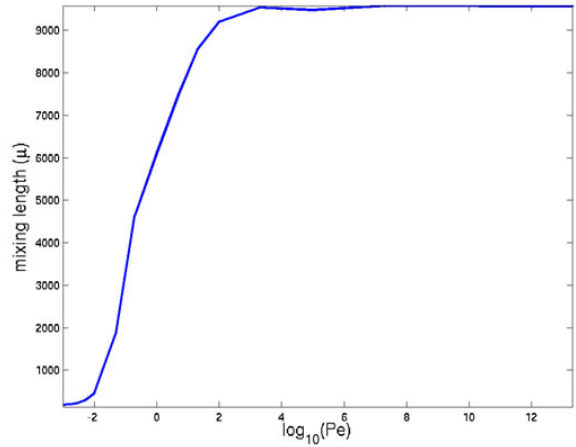

**Figure 6:** The Particle Injection feature in Fluent$^{TM}$ allows for interpolation of velocities to track particles even beyond the diffusive limit of resolution. Here, particles are seen exploiting the ditch to transit to the opposite side of the channel, setting up a pattern of "ditch mixing" which combines alternating shears.

**Figure 7:**
**A**: The logarithm of the standard deviation of species density, here plotted for several values of the Péclet number, falls linearly in the distance. The slope of this line is used to calculate the mixing length (B).
**B:** As expected, the curve flattens near Pe ~ 10$^2$, the limit resolvable at the resolution of the spatial grid used. The lower end is difficult to resolve since only a few, or even two, points can be calculated in (A) before σ is unresolvable.

### IV. Pressure Drop

An additional advantage of CFD is the ability to study pressure drop over the length of one cycle of the pattern, which limits the velocity of the flow and therefore the time needed for effective mixing. On dimensional grounds, we expect the pressure drop, integrated over the surface area, to be approximately $LU\mu \approx 10^{-5}$ N for a channel of length $L$ containing fluid of viscosity $\mu$ which moves at velocity $U$.

This pressure drop is lower for geometries with more grooves since the no-slip condition is effectively weakened over these grooves. One may think of, for example, the difference in drag between a rising bubble and a sedimenting sphere of equal velocity. The pressure drop for the different geometries is listed in Table I and agrees with this intuition.

### V. Dependence on Péclet number

In order to quantify the role of diffusion in the mixing, and hoping to reproduce the mixing vs. Péclet behavior in Figure 3D of [1], we studied variation in σ as a function of span-wise distance for several diffusion constants. Plots of log ($\sigma$) vs. distance do indeed show linear behavior, as expected (cf. Figure 7A). For low Péclet, σ drops rapidly to unresolvable levels as mixing occurs over shorter and shorter lengths. As mentioned in section *II. Mixing*, we can only hope to resolve up to a certain Péclet number (estimated to be ~$10^2$). This trend is evidenced in Figure 7B where we plot the mixing length (obtained from the slopes in Figure 7A) vs. Péclet. Beyond Péclet ~$10^2$, the curve flattens out, as expected.

## VI. Conclusions and Extensions

While CFD is clearly useful for illuminating the details of chaotic mixing flows, a more exciting application would be optimization. In particular, a systematic study over many parameter values (groove depth, width, spacing, number per cycle, etc…) could be of great benefit both to those interested in building better mixers and those interested in understanding how and why they work. An even more appealing numerical study would be an automated iterative parametric optimization, using Fluent[TM] as a "black box" to calculate and return mixing length (or any other appropriate mixing metric) to an optimization routine suited for multi-dimensional parameter space searching, e.g., a Nelder-Mead simplex method. One may also try other experimentally plausible geometries, limited only by the capabilities of current micropatterning.

## VII. Acknowledgements

We gratefully acknowledge Igor Mezic, Howard Stone, Abe Stroock, and Jean-Luc Thieffault for useful discussions; Marc Horner of Fluent[TM] for technical support; and George Whitesides for communicating specific geometric parameters used in the experiment.

**References**

**[1]** Stroock, A.D., Dertinger, S.K.W., Ajdari, A., Mezic, I., Stone, H.A., and Whitesides, G.M., "Chaotic mixer for microchannels." Science Magazine, 295, 2002.
**[2]** Stroock A.D., Dertinger S.K.W., Whitesides G.M., Ajdari, A. "Patterning flows using grooved surfaces." Analytical Chemistry, 74 (20): 5306-5312, 2002
**[3]** Antonsen, T.M., Fan, Z., Ott, E., and Garcia-Lopez, E. Phys. Fluids 8, 3094, 1996.
**[4]** Thiffeault J.L., Childress S. "Chaotic mixing in a torus map." Chaos, 13 (2): 502-507, 2003

|  | **Simple Grooveless Channel** | **Short Legs Removed** | **Original Geometry** |
|---|---|---|---|
| *Inlet* | 1.202e-5 | 8.480e-6 | 8.430e-6 |
| *Half Cycle (758 μm)* | 1.430e-6 | 4.167e-6 | 4.290e-6 |
| *Difference* | 1.059e-5 | 4.323e-6 | 4.14e-6 |
| Pressure Drop | 88.10% | 50.98% | 49.11% |

**Table I:** illustrates the pressure difference in newtons, integrated over the inflow area, given various channel geometries. With increasing patterning, the no-slip condition at the bottom is effectively weakened, allowing flow with decreased pressure drop.